\documentclass[journal,article,submit,moreauthors,pdftex,12pt,a4paper]{mdpi} 
\lastpage{x}
\doinum{10.3390/------}
\pubvolume{xx}
\pubyear{2011}
\history{Received: xx / Accepted: xx / Published: xx}

\pdfoutput=1
\usepackage{amssymb,amsmath}
\usepackage{graphicx}
\usepackage{float}

\Title{Fluctuation, Dissipation and the Arrow of Time}
\Author{Michele Campisi$^{\star}$ and Peter H\"anggi}
\address{Institute of Physics, University of Augsburg,
  Universit\"atsstr. 1, D-86135 Augsburg, Germany}
\corres{michele.campisi@physik.uni-augsburg.de}
\abstract{The recent development of the theory of fluctuation
relations has led to new insights into the ever-lasting question
of how irreversible behavior emerges  from time-reversal
symmetric microscopic dynamics. We provide an introduction to
fluctuation relations, examine their relation to
dissipation and discuss their impact on the  arrow of time
question.}

\begin{document}

\section{Introduction}

Irreversibility enters the laws of thermodynamics in two distinct
ways:
\begin{description}
  \item[Equilibrium Principle]  An isolated, macroscopic system which is placed in an
arbitrary initial state within a finite fixed volume will
attain a unique state of equilibrium.
  \item[Second Law (Clausius)] For a non-quasi-static process
occurring in a thermally isolated system, the entropy change
between two equilibrium states is non-negative.
\end{description}
The first of these two principles is the \emph{Equilibrium
Principle} \cite{Brown01SHPMP32}, whereas the second is the
\emph{Second Law of Thermodynamics} in the formulation given by
Clausius
\cite{Campisi05SHPMP05, Campisi10AJP78}.
Very often the Equilibrium Principle is loosely referred to as
the Second Law of  Thermodynamics, thus creating a great
confusion in the literature. So much that  proposing to raise
the Equilibrium Principle to the rank  of one of the fundamental
laws of thermodynamic became necessary \cite{Brown01SHPMP32}.
Indeed it was argued that this Law of Thermodynamics, defining
the very concept of state of equilibrium, is the most fundamental
 of all the Laws of Thermodynamics (which in fact are formulated
in terms of equilibrium states) and for this reason the
nomenclature \emph{Minus-First Law of Thermodynamics} was
proposed for it.

\begin{figure}[H]
\center
\includegraphics[width=.7\linewidth]{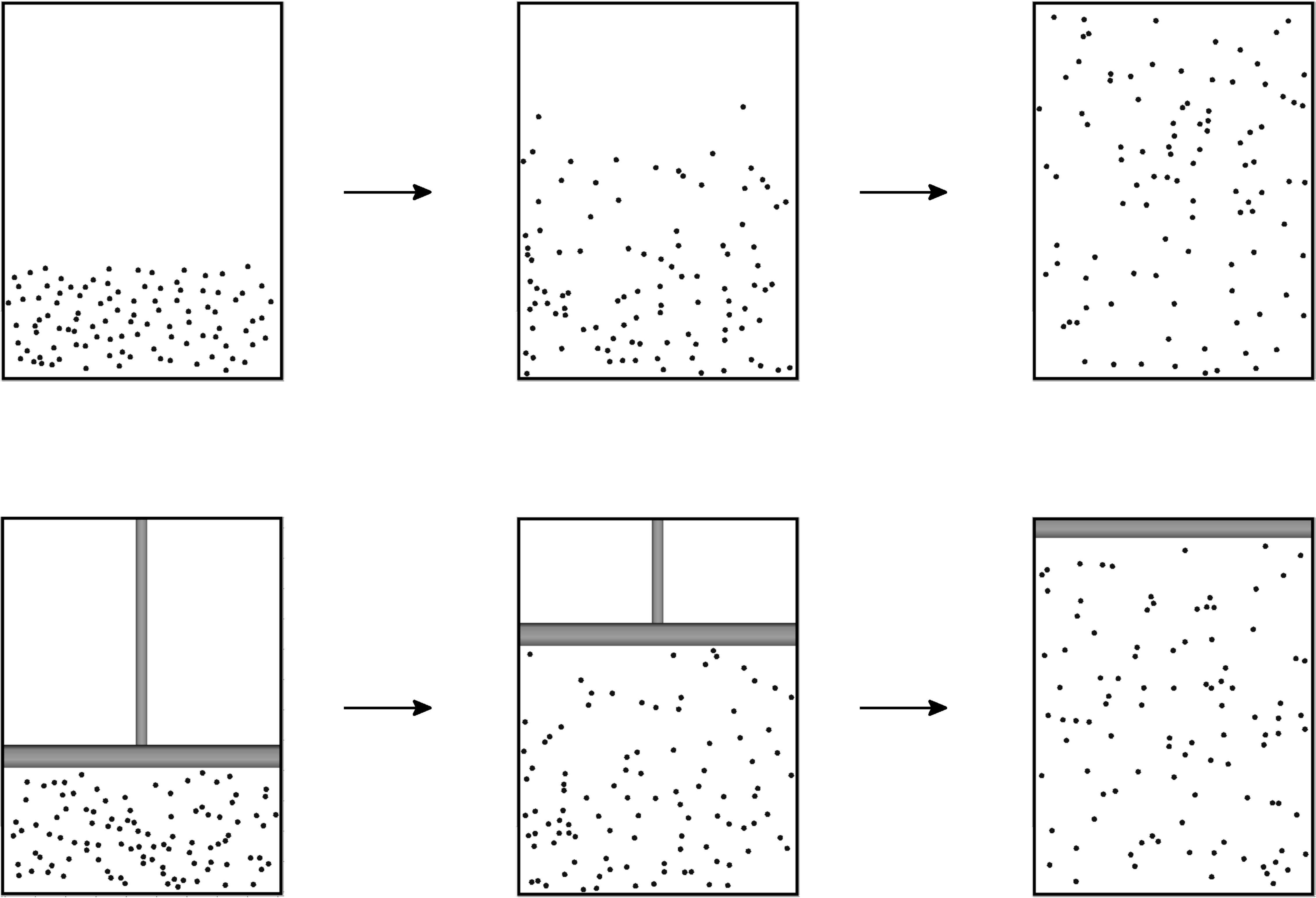}
\caption{Autonomous vs. nonautonomous dynamics. Top:
Autonomous evolution of a gas from a
non-equilibrium state to an equilibrium state (Minus-First Law).
Bottom: Nonautonomous evolution of a thermally isolated gas
between two equilibrium states. The piston moves according
to a pre-determined protocol specifying its position $\lambda_t$
in time.
The entropy change is non-negative (Second Law).
}
\label{fig:microreversibility}
\label{fig:gas}
\end{figure}
The Minus-First Law of Thermodynamics and the Second Law of
Thermodynamics consider two very different situations, see Fig.
\ref{fig:gas}. The  Minus-First Law deals with a completely
isolated system that begins in non-equilibrium and ends in
equilibrium, following its spontaneous and \emph{autonomous}
evolution.
In the Second Law one considers a \emph{thermally} (but not
mechanically) isolated system that begins in equilibrium. A time-dependent
mechanical action perturbs the initial equilibrium, the
action is then turned off and a final equilibrium will be
reached, corresponding to higher entropy.\footnote{That such
final equilibrium state exists is dictated by the Minus-First
Law. Here we see clearly the reason for assigning a higher rank
to the Equilibrium Principle} At variance with the Minus-First
Law, here the system does not evolve autonomously, but rather in
response to a driving: we speak in this case of
\emph{nonautonomous} evolution.

The use of the qualifiers ``autonomous'' and ``nonautonomous''
reflects here the fact that the set of differential equations describing
the microscopic evolution of the system are autonomous (i.e. they
do not contain time explicitly) in cases of the type depicted in Fig.
\ref{fig:gas}, top, and are nonautonomous (i.e. they contain time
explictely) in cases of the type depicted in Fig. \ref{fig:gas}, bottom.
Accordingly the Hamilton function is time independent in the former
cases and time dependent in the latter ones (see Sec. \ref{sec:FT}
below).

In order to illustrate the necessity of clearly distinguishing between the two
prototypical evolutions depicted in Fig. \ref{fig:gas}, let us analyze
one statement which is often referred to as the second law:
\emph{after the removal of a constraint, a system that is initially in
equilibrium reaches a new equilibrium at higher entropy}
\cite{Callen60Book}.
While, after the removal of the constraint the system evolves
autonomously (hence, in accordance to the equilibrium principle
will eventually reach a unique equilibrium state), it is often
overlooked the fact that the overall process is nevertheless described by a set of
nonautonomous differential equations (because the removal of the
constraint is an instance of a external time-dependent mechanical
intervention) with the constrained equilibrium as initial state. Then,
in accordance with Clausius principle the final state is of higher or
same entropy. Thus, this formulation of the second law can be seen
as a special case of Clausius formulation that considers only those
external interventions which are called constraint removals.

Both the Minus-First Law and the Second Law have to do with
irreversibility and the arrow of time. While since the seminal
works of Boltzmann, the main efforts of those working in the
foundations of statistical mechanics were directed to reconcile
the Minus-First Law with the time-reversal symmetric microscopic
dynamics, recent developments in the theory of fluctuation
relations, have brought new and deep insights into the
microscopic foundations of the Second Law. As we shall see below,
fluctuation theorems highlight in a most clear way the
fascinating fact that the Second Law is deeply rooted in the
time-reversal symmetric nature of the microscopic  laws of
microscopic dynamics \cite{Campisi11RMP83, Jarzynski11ARCMP2}.

This connection is best seen if one considers the Second Law in
the formulation given by Kelvin, which is equivalent to Clausius
formulation \cite{Allahverdyan02PHYSA305}:

\begin{description}
  \item[Second Law (Kelvin)] No work can be extracted from a
closed equilibrium system during a cyclic variation of a
parameter by an external source.
\end{description}

The field of fluctuation theorems has recently gained much
attention. Many fluctuation theorems have been reported in the
literature, referring to different scenarios.
Fluctuation theorems exist  for classical dynamics, stochastic
dynamics, and for quantum dynamics; for transiently driven
systems, as well as for non equilibrium steady states;
for systems prepared in canonical, micro-canonical,
grand-canonical ensembles, and even for systems initially in contact
with ``finite heat baths'' \cite{Campisi09PRE80}; they can refer to different
quantities like work (different kinds), entropy production,
exchanged heat, exchanged charge, and even information,
depending on different set-ups. All these developments including
discussions of the experimental applications of fluctuation
theorems, have been summarized in a number of reviews
\cite{Campisi11RMP83, Jarzynski11ARCMP2, Esposito09RMP81,
 Seifert08EJB64}.

In Sec. \ref{sec:FT} we will give a brief introduction to the
classical work Fluctuation Theorem of
Bochkov and Kuzovlev \cite{Bochkov77SPJETP45}, which is
the first fluctuation theorem reported in the literature. The
discussion of this theorem suffices for our purpose of
highlighting the impact of fluctuation theory on dissipation
(Sec. \ref{sec:dissipation}) and on the arrow of time issue
(Sec. \ref{sec:arrow}). Remarks of the origin of time's arrow
in this context are collected in Sec. (\ref{sec:time})

\section{The fluctuation theorem}
\label{sec:FT}

\subsection{Autonomous dynamics}
Consider a completely isolated mechanical system composed of $f$
degrees of freedom. Its dynamics are dictated by some time independent Hamiltonian
$H(\mathbf{q},\mathbf{p})$, which we assume to be time reversal
symmetric; i.e.,
\begin{linenomath}\begin{equation}
H(\mathbf{q},\mathbf{p})=H(\mathbf{q},-\mathbf{p})
\end{equation}\end{linenomath}
Here $(\mathbf{q},\mathbf{p})=(q_1 \dots q_f,p_1\dots p_f)$
denotes the conjugate pairs of coordinates and momenta describing
the microscopic state of the system.

The assumption of time-reversal symmetry implies that if
$[\mathbf{q}(t),\mathbf{p}(t)]$ is a solution of Hamilton
equations of motion, then, for any $\tau$,
$[\mathbf{q}(\tau-t),-\mathbf{p}(\tau-t)]$ is also a solution of
Hamilton equations of motion. This is the well known principle of
\emph{microreversibility} for \emph{autonomous} systems
\cite{Messiah62Book}.

We assume that the system is at equilibrium described by the
Gibbs ensemble:
\begin{linenomath}\begin{equation}
\varrho(\mathbf{q},\mathbf{p})=e^{-\beta
H(\mathbf{q},\mathbf{p})}/Z(\beta)
\label{eq:gibbs}
\end{equation}\end{linenomath}
where $Z(\beta)=\int d\mathbf{p}d\mathbf{q} e^{-\beta
H(\mathbf{q},\mathbf{p})} $ is the canonical partition
function, and $\beta^{-1}=k_B T$, with $k_B$ being the Boltzmann
constant and $T$ denotes the temperature.

We next imagine to be able to observe the time evolution of all
coordinates and momenta within some time span $t \in [0,\tau]$.
Fluctuation theorems are concerned with the
probability\footnote{To be more precise, the probability density
functional (PDFL)}
$P[\Gamma]$ that the trajectory $\Gamma$ is observed. We will
reserve the symbol $\Gamma$ to denote the whole trajectory (that
is, mathematically speaking, to denote a map from the interval
$[0,\tau]$ to the $2f$ dimensional phase space),
whereas the symbol $\Gamma_t$ will be used to denote the specific
point in phase space visited by the trajectory $\Gamma$ at time
$t$.
The central question is how the probability $P[\Gamma]$ compares
with the probability $P[\widetilde{\Gamma}]$ to observe
$\widetilde{\Gamma}$, the time-reversal companion of $\Gamma$:
$\widetilde{\Gamma}_t=\varepsilon\Gamma_{\tau-t}$ where
$\varepsilon(\mathbf{q},\mathbf{p})=(\mathbf{q},-\mathbf{p})$
denotes the time reversal operator.
The answer is given by the microreversibility principle which
implies:
\begin{linenomath}
\begin{equation}
P[\Gamma]=P[\widetilde{\Gamma}] \label{eq:P[Gamma]=P[GammaTilde]}
.
\end{equation}\end{linenomath}

To see this, consider the  Hamiltonian dynamics but for the case that the trajectory $\Gamma$
is {\it not} a solution of Hamilton equations, then $\widetilde
\Gamma$ is also not a solution, and both the probabilities
$P[\Gamma]$ and $P[\widetilde{\Gamma}]$ are trivially zero.
Now consider the case when $\Gamma$ is solution of Hamilton
equations, then also $\widetilde{\Gamma}$ is a solution.
Since the dynamics are Hamiltonian, there is one and only one
solution passing through the point  $\Gamma_0$ at time $t=0$,
then the probability $P[\Gamma]$ is given by the probability to
observe the system at $\Gamma_0$ at $t=0$. By our equilibrium
assumption this is given by $\varrho(\Gamma_0)$ \footnote{To be
more precise $P[\Gamma]\mathcal{D}\Gamma=\varrho(\Gamma_0)
\mathrm{d}\Gamma_0$ where $\mathcal{D}\Gamma$ is the measure on
the $\Gamma$-trajectory space, and $\mathrm{d}\Gamma_0$ is the
measure in phase space}.
Likewise the  $P[\widetilde \Gamma]$  is given by
$\varrho(\widetilde \Gamma_0)$.
Due to time-reversal symmetry and energy conservation we have
$H(\widetilde \Gamma_0)=H(\varepsilon
\Gamma_\tau)=H(\Gamma_\tau)=H(\Gamma_0)$ implying
$\varrho(\widetilde \Gamma_0)= \varrho(\Gamma_0)$,
hence Eq. (\ref{eq:P[Gamma]=P[GammaTilde]}).

To summarize, the micro reversibility principle for autonomous
systems in conjunction with the hypothesis of Gibbsian
equilibrium implies that the probability to observe a trajectory
and its time-reversal companion are equal. There is no way to
distinguish between past and future in an autonomous system at
equilibrium.
Obviously, this is no longer so when the system is prepared out
of equilibrium, as in Fig \ref{fig:gas}, top.

\subsection{Nonautonomous dynamics}
Imagine now the nonautonomous case of a thermally insulated
system driven through the variation of a parameter $\lambda_t$.
Thermal insulation guarantees that the dynamics are still
Hamiltonian. At variance with the autonomous case though, now the
Hamiltonian is time dependent. Without loss of generality we
assume that the varying parameter, denoted by $\lambda_t$
couples linearly to some system observable
$Q(\mathbf{q},\mathbf{p})$, so that the Hamiltonian reads:
\begin{linenomath}\begin{equation}
H(\mathbf{q},\mathbf{p};\lambda_t)=H_0(\mathbf{q},\mathbf{p}
)-\lambda_t Q(\mathbf{q},\mathbf{p})
\label{eq:H(t)}
\end{equation}\end{linenomath}
This is the traditional form employed in the study of the
fluctuation-dissipation theorem \cite{Kubo57aJPSJ12}.\footnote{
For the sake of clarity we remark that the Hamiltonian
describing the expansion of a gas, as depicted in Fig. \ref{fig:gas},
bottom, is not of this form. Our arguments however can be
generalized to nonlinear couplings \cite{Bochkov77SPJETP45}.}
In the following we shall reserve the symbol $\lambda$
(without subscript) to denote the whole parameter variation
protocol, and use the symbol $\lambda_t$, to denote the specific
value taken by the parameter at time $t$. The succession of
parameter values is assumed to be pre-specified (the system
evolution does not affect the parameter evolution).

We assume that $\lambda_t=0$ for $t=0$
and that the system is prepared at $t=0$ in the equilibrium
Gibbs state
\begin{linenomath}\begin{equation}
\varrho_0(\mathbf{q},\mathbf{p})=e^{-\beta H_0(\mathbf{q},\mathbf{p})}/Z_0(\beta)
\, ,
\label{eq:Gibbs}
\end{equation}\end{linenomath}
where
$Z_0(\beta)= \int \mathrm{d}\mathbf{q} \mathrm{d}\mathbf{q}
e^{-\beta H_0(\mathbf{q},\mathbf{p})}$. We further assume that at any {\it fixed} value of the
parameter the Hamiltonian is time reversal symmetric:
\begin{linenomath}\begin{equation}
H(\mathbf{q},\mathbf{p};\lambda_t)=H(\mathbf{q},-\mathbf{p}
;\lambda_t)
\end{equation}\end{linenomath}

Note here the  fact that energy is not conserved in the
nonautonomous case because the Hamiltonian is
time-dependent in this case.
Microreveresibility, as we have described it above, also does not
hold: Given a protocol $\lambda$, if $\Gamma$ is a solution of
the Hamilton equations of motion, in general $\widetilde \Gamma$
is not.
However $\widetilde \Gamma$ is a solution of the equations of
motion generated by the time-reversed protocol $\widetilde
\lambda$, where $\widetilde \lambda_t=\lambda_{\tau-t}$.
This is the \emph{microreversibility principle for nonautonmous
systems} \cite{Campisi11RMP83}. It is illustrated in Fig.
\ref{fig:microreversibility}.
Despite its importance we are not aware of any text-books in
classical (or quantum) mechanics that discusses it.
A classical proof appears in \cite[Sec. 1.2.3]{Stratonovich94Book}.
Corresponding quantum proofs were given in Refs.
\cite{Andrieux08PRL100} and
 \cite[See appendix B]{Campisi11RMP83}.

\begin{figure}[t]
\center
\includegraphics[width=8.5cm]{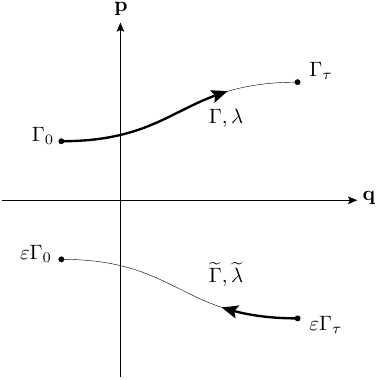}
\caption{Microreversibility for nonautonomous classical
(Hamiltonian) systems. The initial condition $\Gamma_0$ evolves
to $\Gamma_\tau$
under the protocol $\lambda$, following the path $\Gamma$.
The time-reversed final condition
$\varepsilon \Gamma_\tau$ evolves to the time-reversed initial
condition
$\varepsilon \Gamma_0$ under the protocol $\widetilde \lambda$,
following the
path $\widetilde \Gamma$.} \label{fig:microreversibility}
\end{figure}

As with the autonomous case we can ask how the
probability distribution $P[\Gamma,\lambda]$ that the trajectory
$\Gamma$ is realized under the protocol $\lambda$, compares with
the probability distribution $P[\widetilde \Gamma,\widetilde
\lambda]$ that the reversed trajectory $\widetilde\Gamma$ is
realized under the reversed protocol $\widetilde \lambda$. The
answer to this was first given by Bochkov and Kuzovlev
\cite{Bochkov77SPJETP45}, who showed that
\begin{linenomath}
\begin{equation}
P[\Gamma,\lambda]=P[{\widetilde \Gamma,\widetilde \lambda}]
e^{\beta W_0} \label{eq:FT}
\end{equation}\end{linenomath}
where
\begin{equation}
W_0=\int_0^\tau dt \lambda_t \dot Q_t \label{eq:W0} \, .
\end{equation}
Here, $Q_t=Q(\Gamma_t)$ denotes the evolution of the quantity $Q$ along
the trajectory $\Gamma$
and $W_0$ is the so called  ``exclusive work''.
As discussed in
\cite{Jarzynski07CRPHYS8, Horowitz07JSM07, Campisi11RMP83, Campisi11PTRSA369}
yet another definition of work is possible,
the so called  ``inclusive work'' $W=-\int dt \dot\lambda_t Q_t$,
leading to a different and equally important fluctuation theorem
involving  free energy differences
\cite{Campisi11RMP83, Jarzynski97PRL78, Crooks99PRE60}.
Without entering the
question about the physical meaning of the two quantities $W$
and $W_0$, it suffices for the present propose to notice that for
a cyclic transformation $W_0=W$.\footnote{For a detailed discussion on the
differences between the two work expressions we refer the readers to Sect. III.
A in the colloquium \cite{Campisi11RMP83}.}
In the remaining of this section we will restrict our analysis to
cyclic transformations ($\lambda_0=\lambda_\tau$) in order
to make contact with Kelvin postulate and to avoid any 
ambiguity regarding the usage of the word ``work''.

Just like Eq. (\ref{eq:P[Gamma]=P[GammaTilde]}) constitutes
a direct expression of the principle of microreversibility for
autonomous systems, so is Eq. (\ref{eq:FT}) a direct expression
of the more general principle of microreversibility for nonautonomous systems. Remarkably it expresses the second law in a
most clear and refined way.

In order to see this it is important to realize that the work
$W_0$ is odd under time-reversal. This is so because $W_0$ is
linear in a quantity $\dot Q_t$, which is the time derivative of
an even observable $Q$. The theorem says that the probability to
observe a trajectory corresponding to some work $W_0>0$ under the
driving $\lambda$ is exponentially larger than the probability to
observe the reversed trajectory (corresponding to $-W_0$) under
the driving $\widetilde \lambda$. This provides a statistical
formulation of the second law
\begin{description}
\item[Second Law (Fluctuation Theorem)]
Injecting some amount of energy $W_0$ into a thermally
insulated system at equilibrium at temperature $T$
by the cyclic variation of a parameter, is exponentially
(i.e. by a factor $e^{W_0/(k_B T)}$)
more probable than withdrawing the same amount of energy from it
by the reversed parameter variation.
\end{description}

Multiplying Eq. (\ref{eq:FT}) by $e^{-\beta W_0}$ and integrating
over all $\Gamma$-trajectories, leads to the relation \cite{Bochkov77SPJETP45}:
\begin{linenomath}\begin{equation}
\langle e^{-\beta W_0} \rangle_\lambda =1 \label{eq:BKI} \, .
\end{equation}\end{linenomath}
The subscript $\lambda$ in Eq. (\ref{eq:BKI}) is there to recall
that the average is taken over the trajectories generated by the
protocol $\lambda$. In particular, the notation $\langle \cdot \rangle_\lambda$ denotes an
nonequilibrium average.\footnote{
The nonequilibrium average $\langle \cdot \rangle_{\lambda}$ can
be understood as an average over the work probability density
function $p[W_0;\lambda]$, that is the probability that the energy
$W_0$ is injected in the system during one realization of the driving
protocol. It formally reads\cite{Campisi11RMP83}:
$p[W_0;\lambda]=\int \mathrm{d}\mathbf{q}_0 \mathrm{d}\mathbf{p}
_0 \rho_0(\mathbf{q}_0,\mathbf{p}_0)\delta[W_0-\int_0^\tau
\lambda_t  \dot Q(\mathbf{q}_t,\mathbf{p}_t)] $, where $\delta$
denotes Dirac's delta function, and $(\mathbf{q}_t,\mathbf{p}_t)$ is
the evolved of its initial $ (\mathbf{q}_0,\mathbf{p}_0)$  under the driving protocol
$\lambda$.}
Combining Eq. (\ref{eq:BKI}) with Jensen's inequality, $\langle
\exp(x) \rangle \geq \exp({\langle x \rangle)}$, leads to
\begin{linenomath}\begin{equation}
\langle W_0 \rangle_\lambda \geq 0 \;,
\label{eq:2ndLawKelvin}
\end{equation}\end{linenomath}
which now expresses Kelvin's postulate as a nonequilibrium
inequality \cite{Bochkov77SPJETP45}. The quantum version of the fluctuation theorems by Bochkov and Kuzovlev have been given only recently in Ref. \cite{Campisi11PTRSA369}. This latter reference in addition reports  its microcanonical variant, which applies to the case when the system begins in a state of well defined energy.

\section{Dissipation: Kubo's formula}
\label{sec:dissipation}

Before we continue with the implications of the fluctuation
theorem for the arrow of time question, it is instructive to see
in which way the fluctuation theorem relates to dissipation.

Given the distribution $P[\Gamma,\lambda]$, the distribution
$p[Q,\lambda]$ that a trajectory $Q$ of the observable
$Q(\mathbf{q},\mathbf{p})$ occurs
in the time span $[0,\tau]$, can be formally expressed as:
\begin{linenomath}\begin{equation}
p[Q,\lambda]=\int \mathcal{D}\Gamma  P[\Gamma,\lambda] \delta
(Q-Q[\Gamma])
\end{equation}\end{linenomath}
where $\delta$ denotes Dirac's delta in the $Q$-trajectory space,
the integration is a functional integration over all
$\Gamma$-trajectories, and $Q[\Gamma]$ is defined as
$Q[\Gamma]_t\doteq Q[\Gamma_t]$.

Multiplying Eq. (\ref{eq:P[Gamma]=P[GammaTilde]}) by $e^{-\beta
\int \lambda_s \dot Q_s ds} \delta (Q-Q[\Gamma])$ and integrating
over all $\Gamma$-trajectories, one finds:
 \begin{linenomath}\begin{equation}
p[Q,\lambda]e^{-\beta \int \lambda_s \dot Q_s ds}=p[{\widetilde
Q,\widetilde \lambda}] \;, \label{eq:FT-Q}
\end{equation}\end{linenomath}
where $\widetilde Q$ is the time reversal companion of $Q$:
$\widetilde Q_t= Q_{\tau-t}$.
Now multiplying both sides of Eq. (\ref{eq:FT-Q}) by $Q_\tau$ and
integrating over all $Q$-trajectories, one obtains:
\begin{linenomath}\begin{equation}
\langle Q_\tau e^{-\beta \int \lambda_s \dot Q_s
ds}\rangle_\lambda = \langle \widetilde Q_\tau
\rangle_{\widetilde \lambda}
\label{eq:<Q>}\end{equation}\end{linenomath}
Note that $\langle\widetilde Q_\tau \rangle_{\widetilde
\lambda}=\langle Q_0\rangle_{\widetilde \lambda}$ and that, due
to causality, the value taken by
the observable $Q(\mathbf{q},\mathbf{p})$ at time $t=0$ cannot be
influenced by the subsequent evolution of the protocol
$\widetilde \lambda$. Therefore, the average presents a manifest equilibrium
average; that is to say that it is an average over the initial canonical
equilibrium
$\varrho_0(\mathbf{q},\mathbf{p})$. We denote this equilibrium
average by the symbol $\langle \cdot \rangle$ (with no
subscript). Thus,   Eq. (\ref{eq:<Q>}) reads
\begin{linenomath}\begin{equation}
\langle Q_\tau e^{-\beta \int \lambda_s \dot Q_s
ds}\rangle_\lambda = \langle  Q_0 \rangle
\label{eq:<Q>2}\end{equation}\end{linenomath}
By expanding the exponential in Eq. (\ref{eq:<Q>2}) to first
order in $\lambda$, one obtains:
\begin{linenomath}\begin{align}
\langle Q_\tau \rangle_\lambda -\langle Q_0\rangle &= \beta \left
\langle Q_\tau \int_0^\tau \lambda_s \dot Q_s ds \right
\rangle_\lambda +O(\lambda^2) \, .
\end{align}\end{linenomath}
Since the bracketed expression on the rhs is already $O(\lambda)$
we can replace the non-equilibrium average $\langle \cdot
\rangle_\lambda$ with the equilibrium average $\langle \cdot
\rangle$ on the rhs. Further, since averaging commutes with time
integration one arrives, up to order $O(\lambda^2)$, at:
\begin{linenomath}\begin{align}
\langle Q_\tau \rangle_\lambda -\langle Q_0\rangle & =  \beta
\int_0^\tau   \langle  Q_\tau  \dot Q_s  \rangle \lambda_s ds \;,  \\
& = -\beta \int_0^\tau   \langle  \dot Q_{\tau-s} Q_{0}  \rangle \lambda_s ds \;.
\end{align}\end{linenomath}

In the second line we made use of the time-homogeneous
nature of the equilibrium correlation function.
This is the celebrated Kubo formula \cite{Kubo57aJPSJ12} relating
the non equilibrium linear response of the quantity $Q$ to the
equilibrium correlation function  $\phi(s,\tau)=\langle  Q_\tau
\dot Q_s  \rangle$. As Kubo showed it implies the
fluctuation-dissipation relation \cite{Callen51PR83}, linking,
for example, the mobility of a Brownian particle to its diffusion
coefficient \cite{Einstein26Book}, and the resistance of an
electrical circuit to its thermal noise
\cite{Johnson28PR32, Nyquist28PR32}

This classical derivation of Kubo's formula from the fluctuation
theorem is a simplified version of the derivation given by Bochkov
and Kuzovlev \cite{Bochkov77SPJETP45}. The corresponding
quantum derivation was reported by Andrieux and Gaspard
\cite{Andrieux08PRL100}.

\section{Implications for the arrow of time question}
\label{sec:arrow}

Jarzynski has analyzed in a transparent way
how the fluctuation theorem for the inclusive work, $W$,
may be employed to make guesses about
the direction of time's arrow \cite{Jarzynski11ARCMP2}. Here we
adapt his reasoning  to the case of the {\it exclusive}
work, $W_0$, which appears in the fluctuation relation of Bochkov and Kuzovlev,
Eq. (\ref{eq:FT}).

Just imagine we are shown a movie of an experiment in which a system
starting at temperature $T=(k_B \beta)^{-1}$
is driven by a protocol, and we are asked to guess whether the
movie is displayed in the same direction as it was filmed or in
the backward direction, knowing that tossing of an unbiased coin decided the
direction of the movie. When the outcome is $+(-)$,
the movie is shown in the same(opposite) direction as it was filmed.
Imagine next that we can infer from the analysis of each single frame $t$
the instantaneous values $\lambda_t$ and $Q_t$ taken by the
parameter and its conjugate observable, respectively. With these
we can evaluate the work $W_0$ for the displayed process using
Eq. (\ref{eq:W0}).
Envision that we find, for the shown movie that $\beta W_0 \gg 1$. If
the film was shown in the ``correct'' direction it means that a
process corresponding to $\beta W_0 \gg 1$ occurred. If the film
was shown backward then it means that a process corresponding to
$\beta W_0 \ll -1$ occurred (recall that $W_0$ is odd under
time-reversal). The fluctuation theorem tells us that the former
case occurs with an overwhelmingly higher probability relative to
the probability of the latter case. Then we can be very much
confident that the film was running in the correct direction.
Likewise if we observe $\beta W_0 \ll -1$, then we can say with
very much confidence the the film depicts the process in the
opposite direction as it happened. Clearly when intermediate
values of $\beta W_0$ are observed we can still make well
informed guesses about the direction of the movie, but with less
confidence. The worst scenario arises when we observe $W_0=0$, in
which case we cannot make any reliable guess. The question then
arises of how to quantify the confidence of our guesses. This is a
typical problem of Bayesian inference. Before we are shown the
movie our degree of belief of the outcome $+$, is given by the
\emph{prior}, $P(+)=1/2$ (likewise, $P(-)=1-P(+)=1/2)$. After we have seen the movie the prior is
updated to the \emph{posterior}, $P(+|W_0)$, which is the degree of
belief that the outcome $+$ occurred, given the observed work
$W_0$. Using Bayes theorem, the posterior is given by
\begin{equation}
P(+|W_0)= \frac{P(W_0|+)}{P(W_0)}P(+)
\label{eq:Bayes}
\end{equation}
where $P(W_0|+)$ is the conditional probability to observe $W_0$
given that $+$ occurred, and $P(W_0)$ is the probability to observe
$W_0$; i.e., $P(W_0)=P(W_0|+)P(+)+P(W_0|-)P(-)$. According to the
fluctuation theorem $P(W_0|+)/P(-W_0|+)=e^{\beta W_0}$ and since
$W_0$ is odd under time reversal, $P(W_0|-)=P(-W_0|+)$. Using
these relations together  with Eq. (\ref{eq:Bayes})
one obtains:
\begin{linenomath}\begin{equation}
P(+|W_0) = \frac{1}{e^{-\beta W_0}+1}
\label{eq:LF}
\end{equation}\end{linenomath}
\begin{figure}[]
\center
\includegraphics[width=8.5cm]{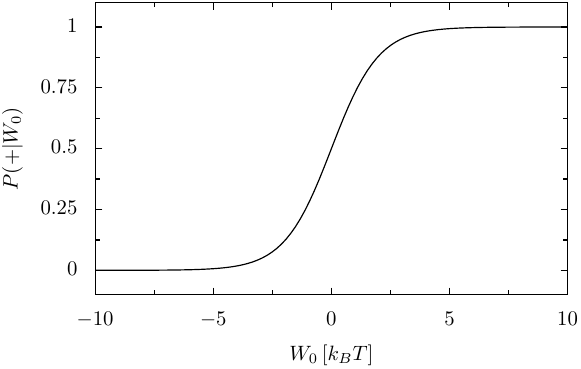}
\caption{Degree of belief $P(+|W_0)$ that a movie showing the nonautonomous
evolution of a system is shown in the same temporal order as
it was filmed, given that the work $W_0$ was observed and that the direction of the movie was decided by the tossing of an unbiased coin. }
\label{fig:likelihood}
\end{figure}
Figure \ref{fig:likelihood} displays $P(+|W_0)$ as a function
of $W_0$.
As it should be $P(+|W_0)$ is larger than $1/2$ for
positive $W_0$, and vice versa, and is an
increasing function of $W_0$. If $W_0$ is large compared to
$\beta^{-1}$, then $P(+|W_0) \simeq 1$, and we can be
almost certain that the movie was shown in the forward direction.
Vice versa, if $\beta W_0 \ll -1$, then we
can say with almost certainty that the movie has been shown
backward.  The transition to certainty of guess occurs quite
rapidly (in fact exponentially) around $|\beta W_0| \simeq 5$.
Note that that for an autonomous system $W_0=0$, implying
$P(+|W_0)=P(-|W_0)=1/2$, meaning that, as we have
elaborated above, there is no way to discern the direction of
time's arrow in an autonomous system at equilibrium.

Since the fluctuation theorem (\ref{eq:FT}) holds
as a general law \emph{regardless of the size of the system},
it appears that our ability to discern the direction of time's
arrow does not depend on the system size.
It is also worth mentioning the role played by thermal
fluctuations in shaping our guesses. Particularly,
with a given observed value $W_0$, the lower the temperature,
the higher is the confidence (and vice-versa).

\section{Remarks}
\label{sec:time}
It emerges from our discussion regarding the arrow of time (Sec.
\ref{sec:arrow}), that the statistical character of the Second Law
becomes visible when the energies injected in a system, $W_0$, are
of the same order of magnitude as the thermal fluctuations, $k_B T$,
regardless of the system size. This means, that,
in contrast to what is sometimes believed, work fluctuations happen
and are experimentally observable in microscopic and
macroscopic systems alike. As a matter of fact, experimental
verifications of the fluctuation theorem have been performed involving
both microscopic systems, e.g. a single macromolecule
\cite{Collin05NAT437, Liphardt02SCIENCE296}, and macroscopic
systems, e.g., a torsional pendulum \cite{Douarche05EPL70}.

As we have mentioned in the introduction, traditionally the question of the
emergence of the arrow of time from microscopic dynamics have been
addressed within the framework of the Minus-First Law. In all existing
approaches the arrow of time emerges from the introduction of some
\emph{extra ingredient} which in turn then dictates the time direction.
Typically, one resorts to a coarse-graining procedure of the microscopic
phase space to describe some state variables. For example, this is so in the
theory of Gibbs and related approaches, see, e.g., in Ref. \cite{Wu1969}. The
time arrow is then generated via the observation that such coarse grained
quantities no longer obey time-reversal symmetric Hamiltonian dynamics.
More frequently, one resorts to additional assumptions which are of a
probabilistic nature: Typical scenarios that come to mind are (i) the use of
Boltzmann Sto{\ss}ahlansatz in the celebrated Boltzmann kinetic theory,  (ii)
the assumption of initial molecular chaos in more general kinetic theories that
are in  the spirit of Bogoliubov, or, likewise, with Fokker-Planck and master
equation dynamics that no longer exhibit an explicit time-reversal invariant
structure \cite{Wu1969, HanggiPR}. All such additional elements then induce
the result of a  \emph{direction in time} with   {\it future} not being equivalent
with  {\it past} any longer.

Having stressed the too often overlooked fact that the Second Law does not
refer to the traditionally considered scenario of autonomously evolving
systems, but rather to the case of nonautonomous dynamics, here we have
focussed on the emergence of time's arrow in a driven system starting at
equilibrium. Having based our derivation on the principle of nonautonomous
microreversibility, Fig. \ref{fig:microreversibility}, the question arises naturally
regarding the origin of the time asymmetry in this case. It originates from the
combination of the following two elements:
i) The introduction of an explicit time dependence of the Hamiltonian, Eq.
(\ref{eq:H(t)}),
ii) The particular shape of the initial equilibrium state, Eq. (\ref{eq:Gibbs}).
The first breaks time homogeneity thus determining the emergence of an
arrow of time, while the second determines its direction. It is in particular the
fact that the initial equilibrium is described by a probability density function
which is a \emph{decreasing} function of energy, that determines the $\geq$
sign in Eq. (\ref{eq:2ndLawKelvin}). An increasing probability density function would
result in the opposite sign
\cite{Allahverdyan02PHYSA305, Campisi08SHPMP39, Campisi08PRE78b}.
In regard to breaking time homogeneity, it is worth commenting that the
assumption of nonautonomous evolution has to be regarded itself as a
convenient and often extremely good \emph{approximation} in which the
evolution $\lambda$ of the external parameter influences the system
dynamics without being influenced minimally by the system.\footnote{ In
principle, one should treat the external parameter itself as a dynamical
coordinate, and consider the autonomous evolution of the extended system.}
This indeed presupposes the intervention of a sort of Maxwell Demon
(i.e., the experimentalist), who predisposes things in such a way that the
wanted protocol actually occurs. This in turn evidences the
phenomenological nature of the Second Law.
It is not a law that dictates how things go by themselves,
but rather how they go in response to particular experimental investigations.

\section*{Acknowledgements}
This work was supported by the cluster of excellence
Nanosystems Initiative Munich (NIM) and the
Volkswagen Foundation (project I/83902).

\bibliographystyle{mdpi}
\makeatletter
\renewcommand\@biblabel[1]{#1. }
\makeatother

\end{document}